# Visible-Spectrum Optical Covert Channels in Commodity Smart Lighting

David Noever, PeopleTec, Inc., Huntsville AL
david.noever@peopletec.com

## Abstract

*Smart light-emitting diode (LED) bulbs are networked devices that produce information-carrying output. This study asks whether that light can carry digital data through changes small enough to be difficult for a person in the room to notice. To explore this covert communication channel, we tested an unmodified smart bulb (Philips WiZ A19) using a camera as the receiver. Binary data were encoded programmatically in nearly matched color and brightness changes. Because the differences were deliberately small across 16 million color steps, the receiver periodically measured known reference states and used them to compensate for changes in the camera and background room illumination. The color method successfully transmitted the short message "hi" using a color difference of $\Delta E = 0.5$, a standard measure of perceptual color distance. The receiver recovered 99% of the transmitted bits, and the complete message passed an integrity check. The brightness method remained decodable at the smallest nonzero integer difference tested, alternating between dimming settings 56 and 54, with 96.2% mean bit accuracy across two runs. When both binary symbols were assigned to the same brightness in a control experiment, accuracy fell to approximately chance. These results show that the visible output of an ordinary smart bulb can serve as a low-rate data channel without hardware modification. Camera color drift can overwhelm small signals unless the receiver is recalibrated during transmission, sudden changes in ambient light can disrupt decoding, and longer messages require stronger error correction. The color experiment operated at a difference below a conventional threshold for human color discrimination. The corresponding perceptual limit for the brightness method has not yet been established experimentally.*

## 1. Introduction

A smart bulb combines two functions that are usually considered separately. It is a networked computer that receives digital commands, and it is a light source that turns those commands into changes in color and brightness. The Philips WiZ A19 used in this study accepts lighting commands from devices on the same local network. This creates a security question that is easy to overlook. Once a network command reaches the bulb, can the emitted light itself become a communication channel?

Previous smart-light security research [1-13] has focused mainly on gaining control of the device. Researchers have examined firmware credentials, companion applications, wireless command injections, and lighting-network protocols [14-15]. Those studies address the path into the bulb. The present work examines the path out of it. Specifically, it asks whether small changes in visible light can encode data that a camera can recover.

The experiment treats the bulb and camera as the two ends of a simple optical link. A transmitter sends ordinary lighting commands over the local network. The bulb converts each command into one of two light states, corresponding to binary 0 or 1. A camera observes the bulb, and software decides which state was present. The central problem is to make the two states far enough apart for the camera to distinguish while keeping the change small from the perspective of a human observer.

Two ways of creating those states are tested. In the color method, 0 and 1 are represented by two nearly identical colors [Figure 1]. In the brightness method, the color remains fixed and 0 and 1 are represented by two nearby brightness settings. Very small color differences create an additional problem because the camera's estimate of color drifts over time. The receiver therefore inserts known reference states throughout the message and repeatedly updates its estimate of what 0 and 1 look like.

A feedback procedure provides a second form of adaptation. It alternates the two candidate states, observes the light measured by the camera, and calculates how far apart those states actually appear. This matters because the bulb, lampshade, camera, and ambient illumination all affect the received signal. Figure 1 summarizes the path from network command to emitted light, camera measurement, and recovered data.

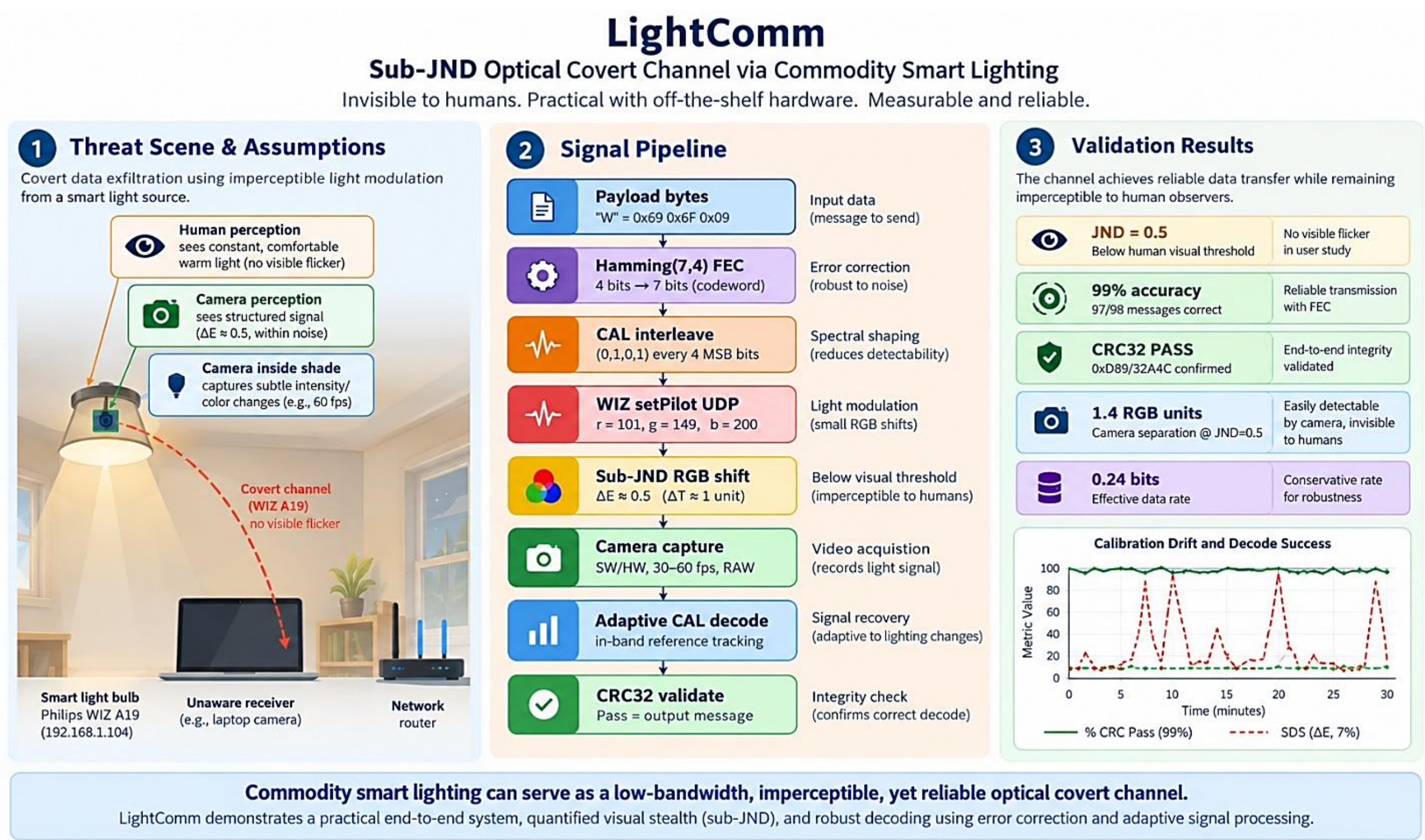

*Figure 1 System overview. Left: Philips WiZ A19 bulb with SVPRO global-shutter camera. Center: WiZ UDP protocol path and network context. Right: six-step Python decode pipeline with validated terminal output.*

## 1.1 Contributions

The experiments establish four novel findings. A short message was recovered through the color method at a color difference of $\Delta E = 0.5$ with 99% raw bit accuracy and a valid end-to-end integrity check. The brightness method remained decodable at the smallest nonzero integer difference tested, settings 56 and 54, with 96.2% mean raw bit accuracy across two runs. Repeated reference measurements corrected a failure caused by camera color drift. A feedback procedure also identified conditions in which the bulb could not reproduce the requested color difference. The experimental record includes failed transmissions and a zero-signal control so that these limits remain visible alongside the successful results.

## 2. Background

As summarized in Table 1, previous work has examined optical information leakage channels. Loughry and Umphress [1] showed that status LEDs on network equipment can unintentionally reveal information about the data being processed, and a later survey documents this class of optical emanations [2]. Later work deliberately controlled lights to create communication channels. xLED used router and switch indicators [3], LED-it-GO used a hard-drive activity light [4], CTRL-ALT-LED used keyboard lock-key LEDs [7], aIR-Jumper used infrared emitters on security cameras [5], and BRIGHTNESS varied a computer display's brightness by amounts intended to remain below visual detection [6]. Table 1 summarizes this progression from unintended optical leakage to deliberately modulated light.

**Table 1. Optical TEMPEST and related covert channel systems.**

| System | Year | Carrier | Rate | Human detection |
|---|---|---|---|---|
| **Loughry and Umphress** | 2002 | Network gear status LEDs | 9600 baud (passive) | Low; passive intercept |
| **Guri et al. xLED** | 2017 | Router and switch status LEDs | 10 to 1,000 bit/s | Visible LED flicker |
| **Guri et al. LED-it-GO** | 2017 | Hard disk activity LED | Up to 4,000 bit/s | Imperceptible above 400 Hz |
| **Guri et al. aIR-Jumper** | 2017 | Security camera IR emitters | ~20 bit/s | Not visible (infrared) |
| **Guri BRIGHTNESS** | 2020 | Screen brightness | Not reported | Not visible (sub-threshold) |
| **Maiti and Jadliwala** | 2019 | Smart bulb IR LEDs | Not reported | Not visible (infrared) |
| **Zhou et al.** | 2019 | Smart LED array | QR hijacking | Not visible to unaided eye |
| **LightComm color axis (this work)** | 2025 | Smart bulb visible | 0.24 data bit/s | Not visible (sub-JND) |
| **LightComm dimming OOK (this work)** | 2025 | Smart bulb dimming | 0.44 data bit/s | Near luminance threshold |

These systems differ in how they gain control of the light source and in whether the signal is visible to a person. Some require software running on the target computer. Others use infrared light, which is outside normal human vision. The present study instead uses the ordinary network interface of a commercial smart bulb and keeps the carrier in the visible spectrum.

The closest predecessor is Light Ears [8], which used the infrared output of a smart bulb as a covert channel. Its authors chose infrared because visible color changes were expected to be noticed. Light Auditor [9] later examined whether electrical power measurements could reveal that infrared transmission. The present study asks whether visible output from the same general class of device can carry data when the changes are made much smaller.

Other work helps explain why a camera can recover changes that may be difficult for a person to notice. Zhou and colleagues [10] used controlled LED flicker to change how a QR code appeared to a camera. Ronen and colleagues [11] demonstrated a self-propagating ZigBee attack in Philips Hue bulbs. That work concerns compromise of the lighting network, while the present study concerns information carried in the emitted light.

The main receiver problem in this study is that the signal is small enough for camera drift to overshadow the signal. The receiver therefore measures known symbols throughout the transmission and continually refreshes its reference. A separate feedback procedure measures the actual optical separation produced by the bulb instead of assuming that a requested numerical change produces the same change at the camera.

## 3. WiZ Protocol Modalities

The Philips WiZ A19 contains a WiZ 8285 1517PT Wi-Fi module based on the ESP8285 microcontroller [12]. Separate pulse-width-modulation (PWM) channels control its red, green, blue, warm-white, and cool-white LED elements. The bulb receives lighting instructions as JavaScript Object Notation (JSON) messages sent with the User Datagram Protocol (UDP) on port 38899. In the tested configuration, these commands did not require authentication. A command can set color, brightness, color temperature, or a predefined lighting scene.

Table 2 lists these controls because each can change the bulb's optical output. Only two are tested as data channels here. The first coordinates the red, green, and blue controls to create two nearby colors. The second changes only brightness. The remaining entries identify possible extensions and are not experimental results of this study.

**Table 2. WiZ protocol parameters as covert channels. Validated column indicates hardware testing in this work.**

| Modality | Parameter | Bits/symbol | Validated | Receiver |
|---|---|---|---|---|
| **Color axis a*** | r, g, b (coordinated) | 1 | Yes | Color camera |
| **Dimming OOK** | dimming (10 to 100) | 6.5 max | Yes | Photodiode or camera |
| **Color temperature** | temp (2200 to 6500 K) | 12 max | No | Spectroradiometer |
| **Scene identity** | sceneId (1 to 33) | 5 | No | Color camera |
| **Scene speed** | speed (10 to 200) | 7.6 | No | Photometer |
| **Power state** | state (boolean) | 1 | No | Any light sensor |
| **Smart plug cycling** | AC outlet state | 1 | No | Any light sensor |

## 4. System Design

### 4.1 Hardware

The transmitter was a Philips WiZ A19 9 W bulb. Its firmware smooths color transitions over roughly half a second. Commands sent substantially faster therefore blend together instead of producing clean, separate optical states. Each transmitted bit was consequently given 750 ms. A typical command is shown below. It specifies the red, green, and blue values together with the dimming setting.

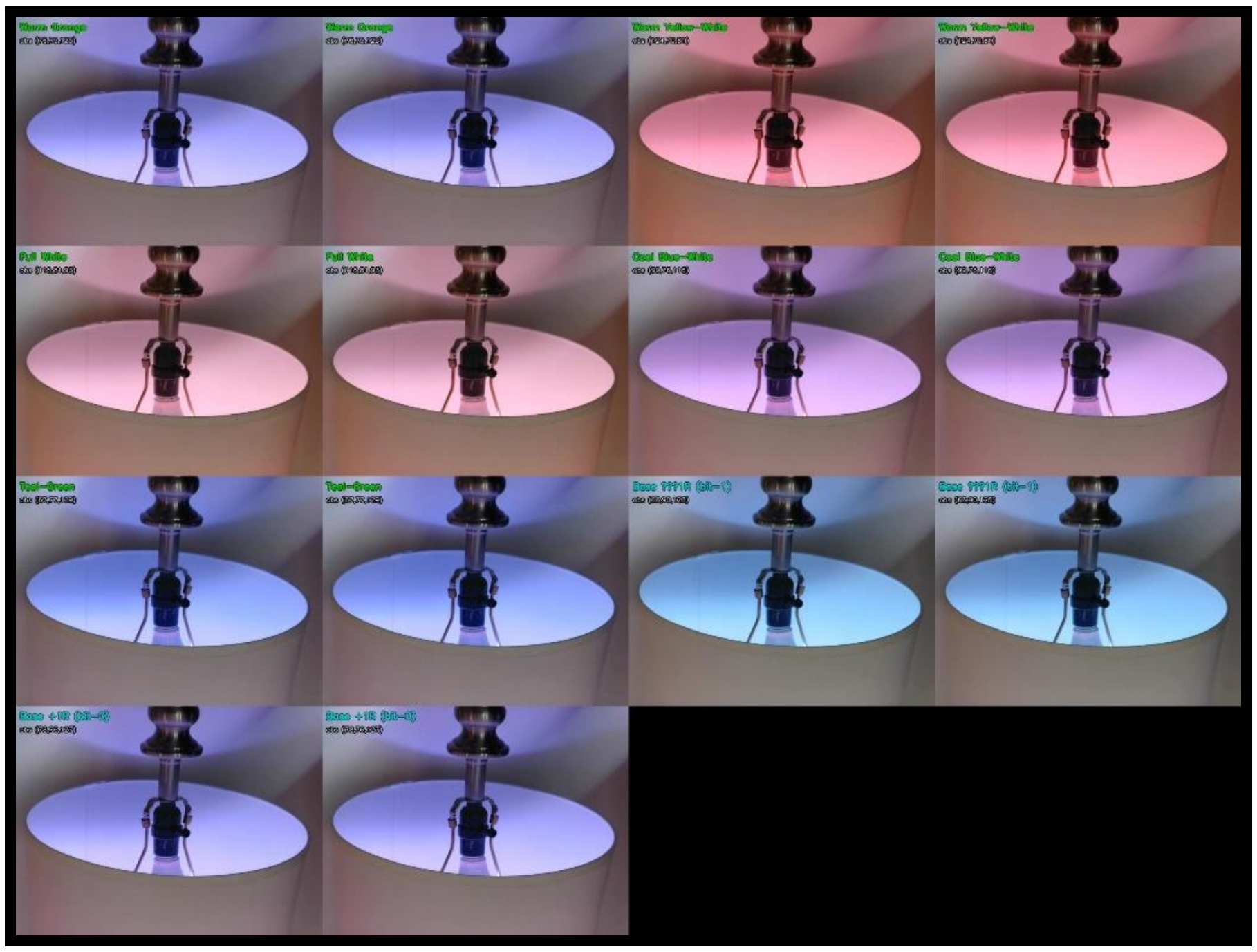

*Figure 2. Color survey grid. The five upper panels show major color variants used for pipeline validation. The bottom-left two panels show bit-0 (Base +1R) and bit-1 (Base -1R) at JND = 0.5; these are visually indistinguishable.*

The receiver was an SVPRO two-megapixel camera with a 5-50 mm f/2.8 zoom lens [13]. It uses a global shutter, meaning that the full image is exposed at essentially the same time. This avoids the row-by-row timing of a rolling-shutter camera, which could mix successive light states near a bit boundary. The camera was placed inside or immediately beside the lampshade and recorded 30 frames each second. For each frame, the software averaged a 60 × 60 pixel region near the center of the bulb image, giving about 22 measurements during each 750 ms bit.

The camera controls above reduce automatic white balance and automatic exposure, though they do not eliminate them. The remaining color drift becomes important when the two transmitted colors are very close. Rather than assume that the first measurement remains valid throughout a message (Appendix A, Figure 3), the receiver periodically measures fresh reference symbols as described in Section 4.4.

### 4.2 Color Axis Channel

The color experiment needs a numerical way to describe two colors that are close in human perception. Raw red, green, and blue values are inconvenient because equal numerical RGB changes do not correspond to equal perceived changes. The experiment therefore uses the CIE L*a*b* color representation. L* represents lightness, a* runs approximately from green to red, and b* runs approximately from blue to yellow. The distance between two colors is summarized here by ΔE. Smaller ΔE values indicate colors that are more difficult to distinguish.

Binary 0 and 1 are produced by moving a small distance in opposite directions along the a* coordinate while keeping the other coordinates as stable as possible. The starting command is RGB (100, 149, 200). The lampshade and camera alter the measured color, so the camera sees approximately RGB (107, 114, 225). This offset is not itself a problem because decoding depends on separating the two received states from one another. At ΔE = 0.5, the two commands differ by only a few integer RGB values after conversion. Figure 2 shows these small symbol states alongside larger color changes used to verify the processing pipeline.

### 4.3 Dimming OOK Channel

The brightness method removes the color calculation. The same base color is used for every bit, and only the bulb's dimming setting changes. One brightness level represents 0 and a nearby level represents 1. The camera measures the mean brightness in the same image region used for the color experiment and compares the observation with the two current reference levels.

*Table 3. Color axis channel experimental results.*

| Message | Bytes | Wire bits | JND | Sep (RGB) | Cal drift | Wire acc | CRC | Result |
|---|---|---|---|---|---|---|---|---|
| **'hi'** | 2 | 214 | 30 | ~122 | < 2.0 | 100% | PASS | PASS |
| **'hi'** | 2 | 214 | 10 | ~31 | < 2.0 | 100% | PASS | PASS |
| **'hi'** | 2 | 214 | 5 | ~16 | < 2.0 | 100% | PASS | PASS |
| **'hi' (v7, fixed cal)** | 2 | 214 | 2 | 5.6 | ~10 over 15 s | 56% | FAIL | FAIL: AWB drift exceeded signal |
| **'hi' (v8)** | 2 | 214 | 1 | 4.0 | 2.2 | 100% | PASS | PASS |
| **'hi' (v8)** | 2 | 214 | 0.5 | 1.4 | 0.7 | 99% (97/98) | PASS | PASS: one bit FEC-corrected |
| **'we attack at dawn'** | 17 | 632 | 0.5 | ~1.2 | 4.7 | 92.5% | FAIL | FAIL: length field corrupted |
| **'SOS Help'** | 8 | 382 | 0.5 | 1.0 to 11.3 | 16.5/20.9 | 79.1% | FAIL | FAIL: ambient step change block 10 |

The size of the brightness change is called the dimming swing, meaning the numerical difference between the two commanded settings. All reported runs are centered on setting 55. A swing of 2 therefore uses 56 for one symbol and 54 for the other. The bulb's command range runs from 10 to 100. This scale is not a direct measure of perceived brightness. A two-unit difference should therefore be read as the smallest nonzero setting tested. It is not proof of a two-percent perceptual change.

### 4.4 Interleaved Recalibration

When the two symbol states are very close, the receiver needs recent examples of both. Each transmission begins with a 16-bit alternating sequence of known 0s and 1s to establish the initial references. After every four payload bits, the transmitter sends the known sequence [0, 1, 0, 1]. These inserted symbols serve as calibration blocks and carry no message data. The receiver measures them and replaces its previous references before decoding the next four payload bits. In the $\Delta E = 0.5$ color run, drift within one six-second calibration interval was about 0.7 RGB units,

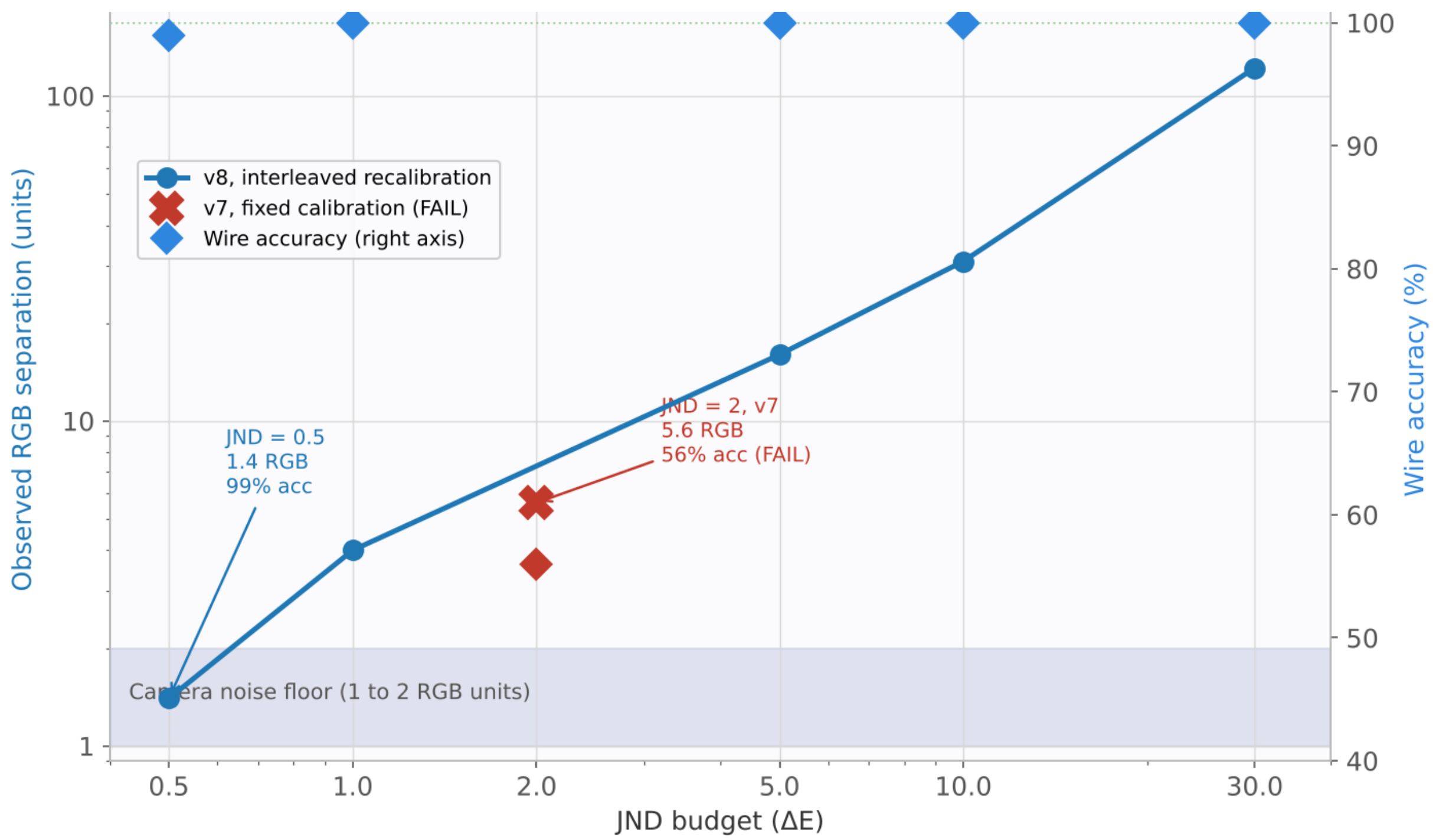

*Figure 3. Observed RGB Signal Separation and Wire Accuracy vs. Perceptual JND Budget (log-log scale; noise floor band shaded; no recalibration). Blue circles: v8 with interleaved recalibration. Red cross: v7 at JND=2 with fixed calibration, 56% accuracy despite separation above the noise floor. Shaded band: camera noise floor at 1 to 2 RGB units*

compared with a nominal symbol separation of about 1.4 RGB units. Appendix A shows the complete frame structure (Figure 9).

### 4.5 Forward Error Correction

Small signals also produce occasional bit errors, so the transmitted message contains redundant information. The color experiments use a Hamming(7,4) error-correcting code. Four data bits become a seven-bit codeword that can repair one erroneous bit. The brightness experiments use threefold repetition, in which each bit is sent three times and the majority value is retained. After error correction, a 32-bit cyclic redundancy check (CRC-32) tests whether the reconstructed message matches the transmitted data. A passing CRC indicates successful recovery of the complete message instead of merely a high percentage of individual bits.

### 4.6 Closed-Loop Optimizer

The feedback procedure measures the channel as the camera actually sees it. It alternates the two candidate states, waits for the bulb to settle, averages several camera frames, and calculates the distance between the measurements. For brightness modulation, it progressively reduces the difference between the two dimming settings. For color modulation, it can also shift the starting RGB command to compensate for a systematic color offset. This matters because a numerical command difference is useful only if it produces a measurable difference in emitted light. The browser dashboard displays the camera feed and measured separation so that loss of separation can be seen during calibration.

## 5. Experimental Results

### 5.1 Color Axis Channel

*Table 4. Selected calibration block measurements, 'hi' at JND = 0.5. 21 of 25 blocks below the 1.0 RGB noise floor. All decoded correctly.*

| Block | ref0 (RGB) | ref1 (RGB) | Sep (RGB) | Note |
|---|---|---|---|---|
| **0** | (103,113,223) | (102,113,223) | 0.78 | below threshold |
| **1** | (104,114,224) | (104,115,225) | 0.86 | below threshold |
| **4** | (106,115,226) | (105,115,225) | 1.12 | above threshold |
| **6** | (106,115,225) | (105,115,225) | 1.03 | above threshold |
| **9** | (107,115,225) | (106,115,225) | 0.65 | below threshold |
| **21** | (107,113,225) | (106,114,225) | 1.04 | above threshold |
| **24** | (107,113,225) | (106,113,225) | 0.97 | below threshold |

**Table 3** is the complete color axis experimental record. Eight transmissions were conducted at six JND levels and three message lengths using the hardware in Section 4.1 with the camera inside the lampshade.

Three experiments at JND = 5, 10, and 30 confirm the software pipeline is correct before approaching the noise floor. At JND = 30 the color change is visually obvious and the observed separation is approximately 122 RGB units. All three achieve 100% wire accuracy with calibration drift below 2 RGB units throughout.

The key comparison is JND = 2 with a fixed calibration reference versus JND = 1 with interleaved recalibration. At JND = 2 the signal separation was 5.6 RGB units, above the camera noise floor, yet wire accuracy was only 56%. AWB drift accumulated approximately 10 RGB units

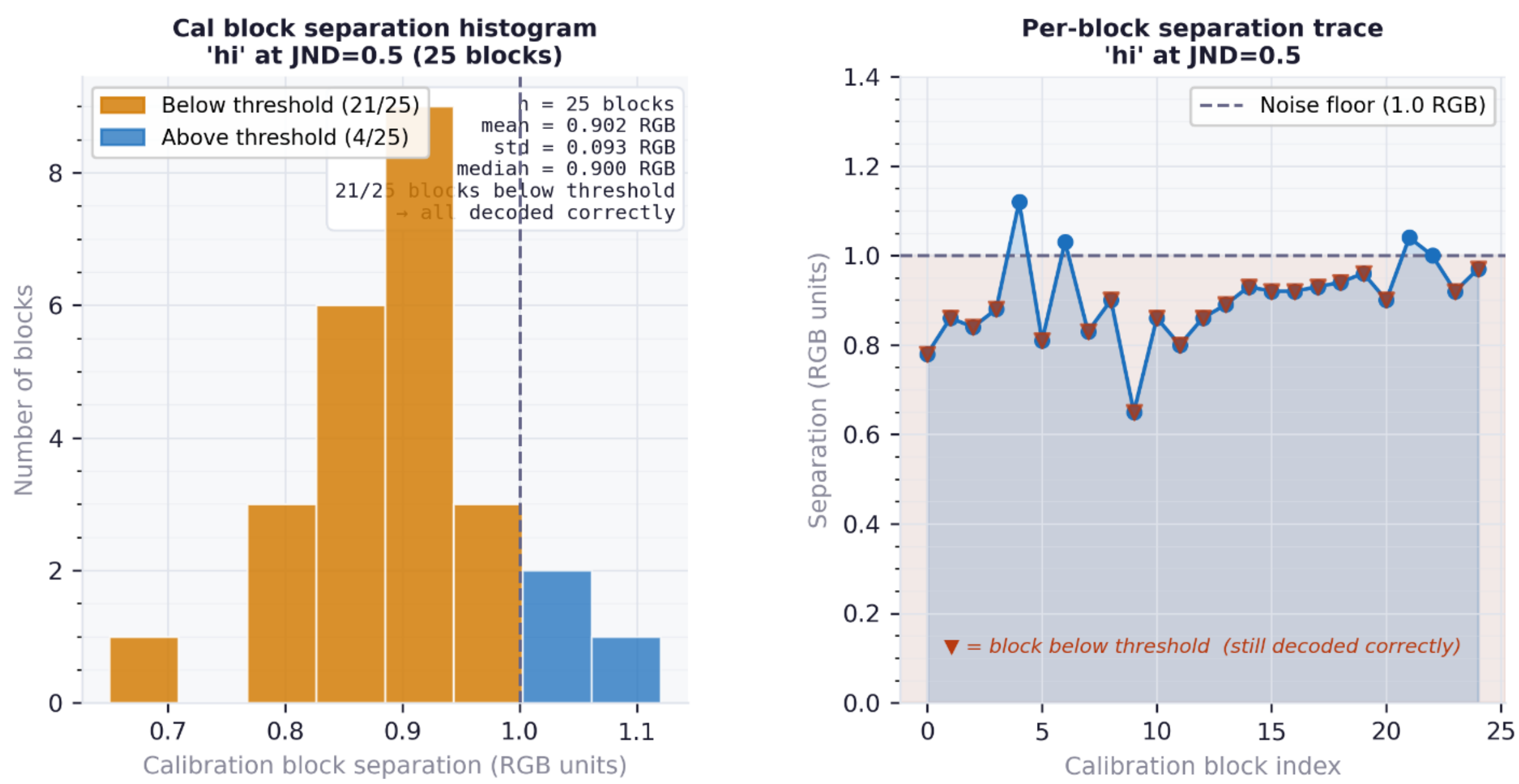

*Figure 4. Calibration block distribution for 'hi' at JND = 0.5. Left: 21 of 25 blocks (amber) below the 1.0 RGB noise floor (mean 0.902, std 0.093). Right: per-block trace; red triangles mark below-threshold blocks, all of which decoded correctly.*

over 160 seconds and displaced the fixed reference past the signal, inverting bit decisions in the second half of the transmission. At JND = 1 with interleaved recalibration, wire accuracy was 100% with 4.0 RGB separation and only 2.2 RGB of drift. Reliability is set by the architecture change. The JND value does not determine it. **Figure 3** plots separation and accuracy

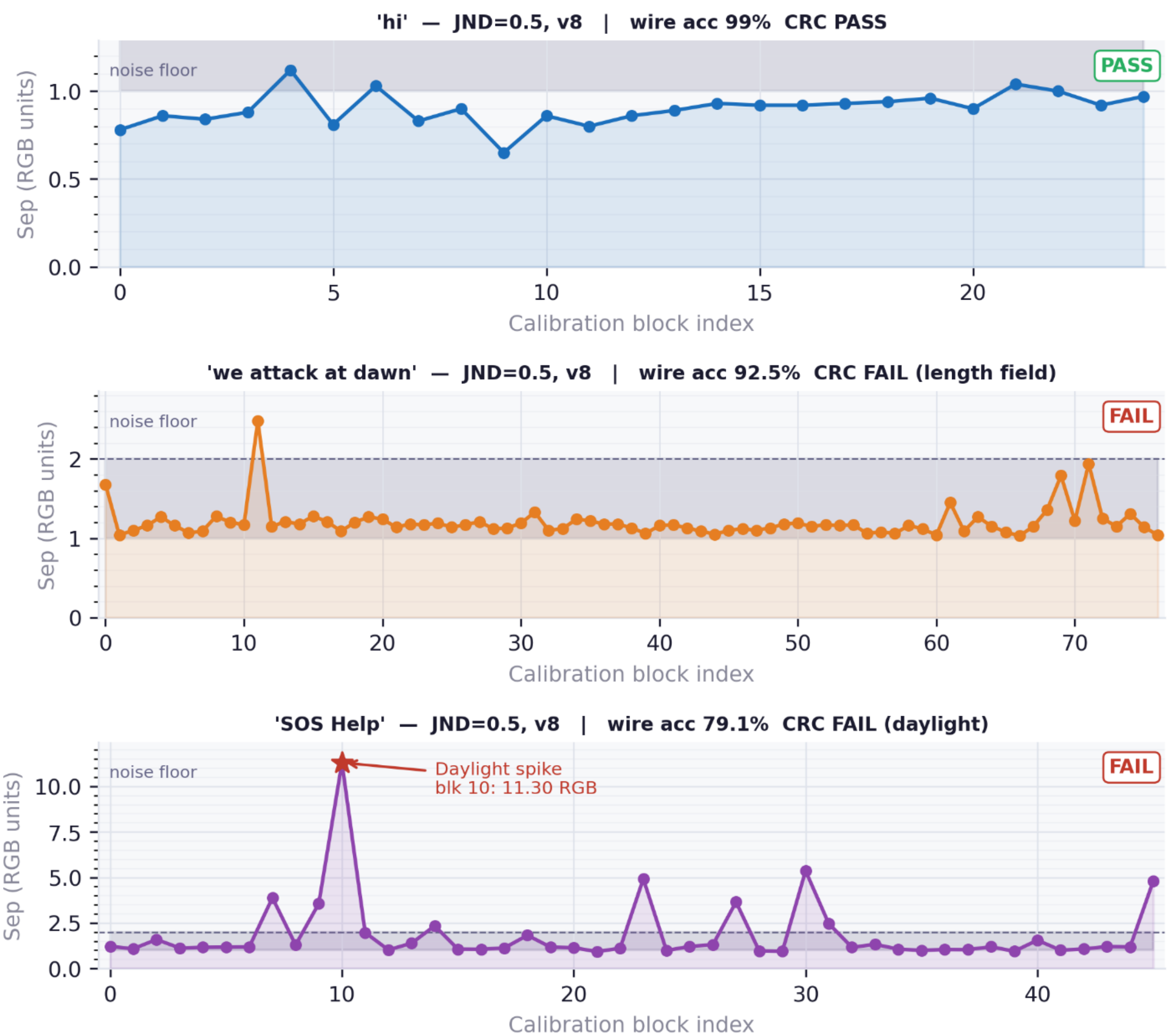

*Figure 5. Per-block calibration separation for the three JND = 0.5 transmissions. Top: 'hi' (PASS), stable throughout. Middle: 'we attack at dawn' (FAIL), normal channel behavior but FEC insufficient. Bottom: 'SOS Help' (FAIL), ambient step change visible at block 10 (11.30 RGB).*

across all JND levels.

The primary result is JND = 0.5 with 99% wire accuracy and exact CRC (0xD8932AAC). Of 25 calibration blocks, 21 reported separations below the 1.0 RGB noise floor. The decoder succeeded because it compares the two symbol states within each calibration window instead of against an absolute reference. The relative ordering is preserved even when absolute discrimination is near impossible. **Table 4** and **Figure 4** show selected block measurements.

Two longer transmissions at JND = 0.5 failed. For 'we attack at dawn' (17 bytes, 474 s), wire accuracy was 92.5% and the decoder aborted with a corrupted length field. At 92.5% accuracy over 44 Hamming codewords, approximately 4 codewords are expected to receive two or more errors, which Hamming (7,4) cannot correct. One of these covered the length field. The calibration data for this run is consistent with normal channel operation. Drift was 4.7 RGB units over 474 seconds and mean block separation was 1.2 RGB units. The cause is the limited strength of the error-correcting code. The channel itself performed normally. Figure 7 relates wire accuracy to message length and shows the Hamming correction limit.

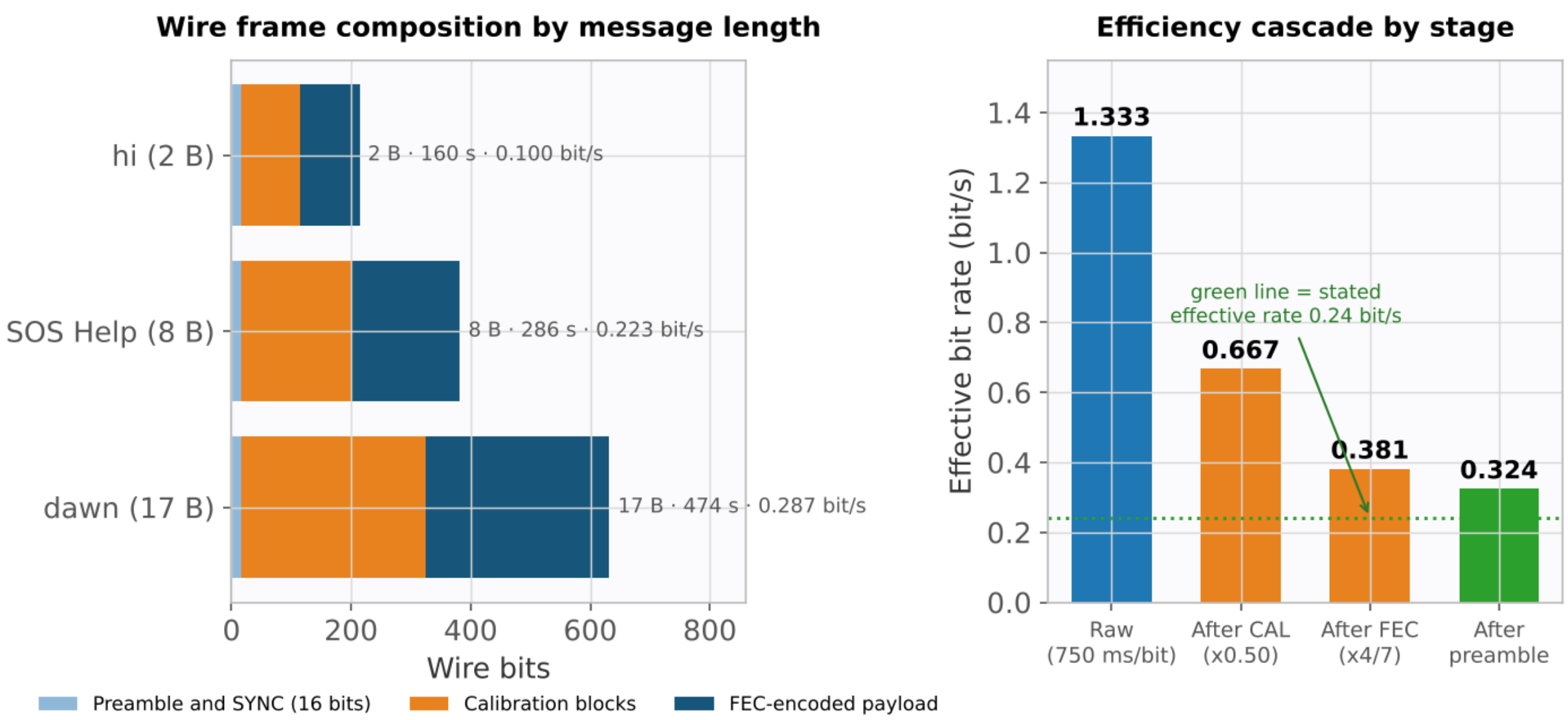

*Figure 6. Throughput breakdown for the color axis channel. Left: wire frame composition by message length. Right: efficiency cascade from raw symbol rate to effective payload rate. CAL overhead (50%) and FEC efficiency (4/7) are the dominant losses.*

For 'SOS Help' (8 bytes, 286.5 s), wire accuracy was 79.1% and block 10 recorded a separation of 11.30 RGB units, eight times the expected value at JND = 0.5. This is consistent with an ambient light step change during that 6-second calibration window. A nearby computer screen turned off during the run, reducing ambient light and changing the room's color balance. The channel would have succeeded in stable lighting conditions**. Figure 5** shows the three per-block calibration traces.

## 5.2 Dimming OOK Channel

**Table 5** presents the complete dimming OOK results across 15 runs. Center dimming was fixed at 55 throughout. All runs used 3x repetition coding, 304 wire bits, and the message 'hi'. The null control at swing=0 uses identical dimming for both symbols and confirms that a 54.9% wire accuracy result represents chance-level performance, ruling out systematic bias in the decoder.

The channel passes at every swing level from 80 down to 2 and fails only at the null control. Swing=2 (dim=56 vs dim=54) achieves 96.2% mean wire accuracy with 2.06 mean RGB separation. Swing=6 and swing=4 were each run twice. Both are consistent at the two-run level. A swing=6 failure in the initial continuous search was traced to a nearby computer screen turning off mid-transmission. A repeat run in isolation gave 97.9% accuracy.

Wire accuracy is flat between 94 and 99% across all tested swings from 80 down to 2, then drops to 54.9% at swing=0. This sharp transition confirms a quantization-limited channel instead of a gradual SNR-limited one. The camera either resolves the dimming difference or it does not, with no intermediate regime. Figure 8 plots observed separation and wire accuracy across the swing range.

**Table 5. Dimming OOK channel, complete swing search (center dim=55, 3x repetition FEC, message='hi').**

| Swing | dim_hi | dim_lo | Runs | Wire acc | Mean sep (RGB) | Result |
|---|---|---|---|---|---|---|
| **80** | 95 | 15 | 1 | 97.9% | 77.27 | PASS |
| **60** | 85 | 25 | 1 | 97.9% | 59.24 | PASS |
| **40** | 75 | 35 | 1 | 98.6% | 37.47 | PASS |
| **30** | 70 | 40 | 1 | 98.6% | 27.88 | PASS |
| **20** | 65 | 45 | 1 | 96.5% | 18.62 | PASS |
| **14** | 62 | 48 | 1 | 95.8% | 14.61 | PASS |
| **10** | 60 | 50 | 1 | 95.8% | 9.34 | PASS |
| **8** | 59 | 51 | 2 | 97.9% | 7.19 | PASS |
| **6** | 58 | 52 | 2 | 97.9% | 5.88 | PASS |
| **4** | 57 | 53 | 2 | 95.1% | 3.72 | PASS |
| **2** | 56 | 54 | 2 | 96.2% | 2.06 | PASS |
| **0 (null control)** | 55 | 55 | 1 | 54.9% | 0.03 | FAIL |

# 6. Discussion

## 6.1 Comparison with Prior Work

The Guri et al. air-gap channels achieve substantially higher throughput. LED-it-GO reaches 4,000 bit/s by scheduling disk I/O on the monitored machine. xLED reaches 1,000 bit/s by controlling router status LEDs via malware. These systems require code execution on the target host. LightComm does not. The attacker needs only WiFi adjacency to the bulb's subnet. The threat models also differ. Guri-style channels exfiltrate from already-compromised hosts, while LightComm creates a channel from the network without touching the monitored machine.

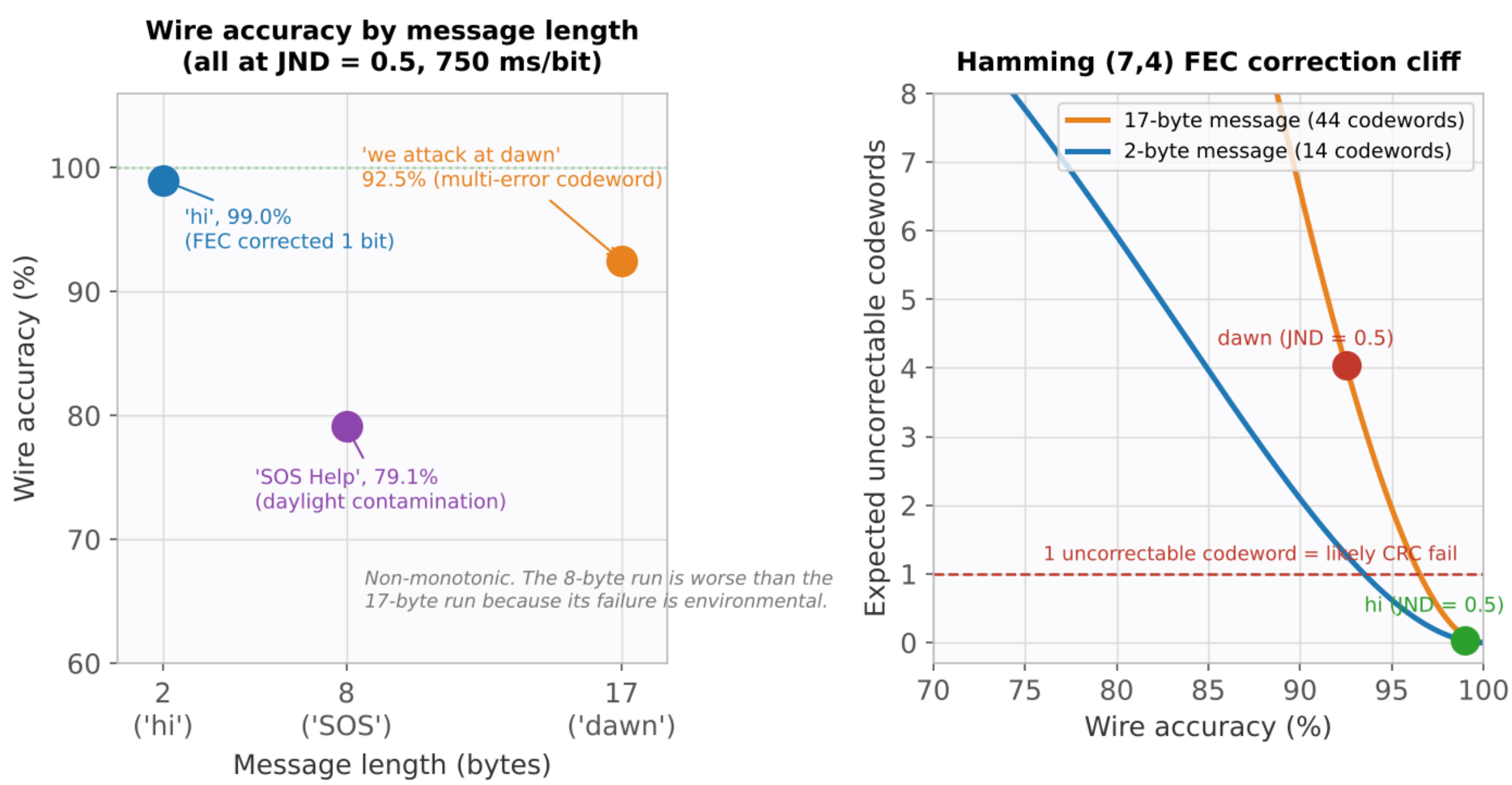

*Figure 7. Failure analysis. Left: non-monotonic accuracy versus message length; the 8-byte run is worse than the 17-byte run because its failure is environmental. Right: Hamming(7,4) FEC correction cliff at 92.5% accuracy over 44 codewords.*

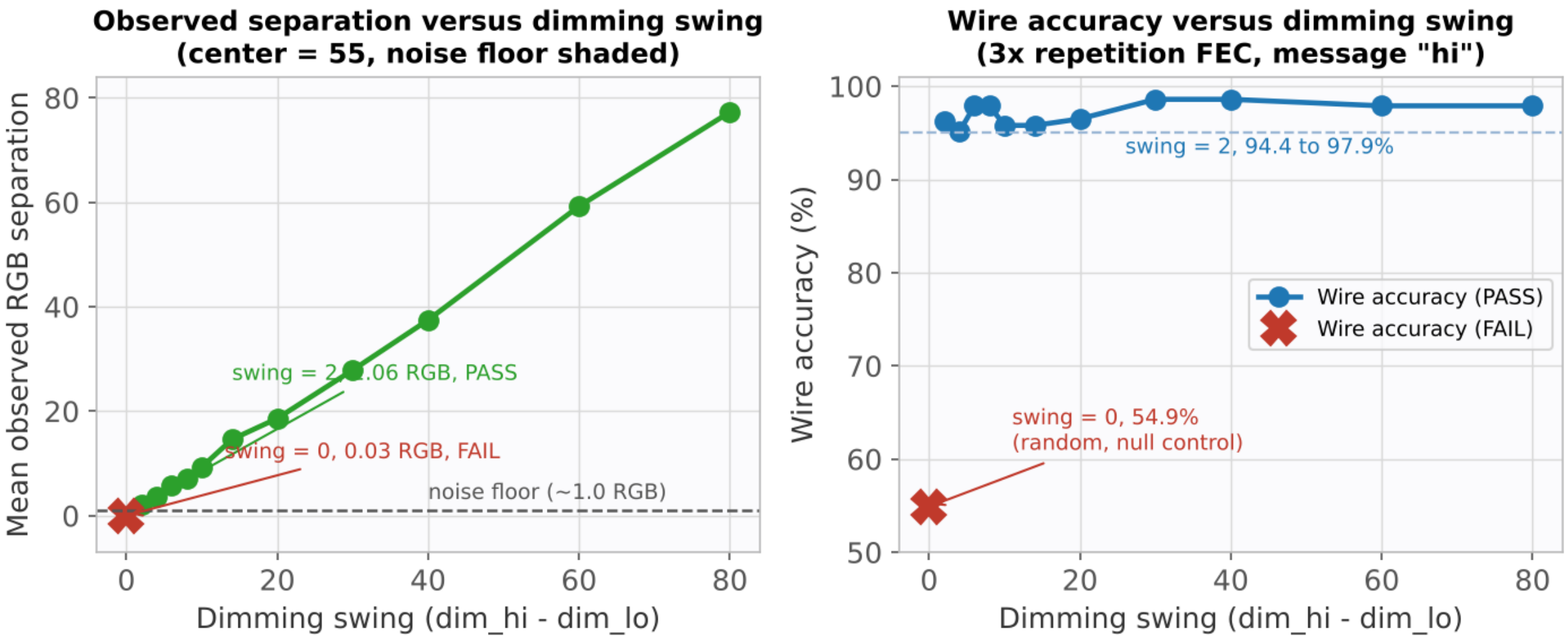

*Figure 8. Dimming OOK swing search. Left: observed RGB separation versus swing; the linear relationship (slope 0.96 RGB units per dimming unit) confirms the camera is resolving the dimming modulation directly. Dashed line: 1.0 RGB noise floor. Right: wire accuracy across the operating range.*

BRIGHTNESS (Guri, 2020) is the closest prior work in terms of operating below the visual threshold. It modulates screen brightness at sub-perceptual amplitudes and, like LightComm, operates in the visible spectrum without requiring IR hardware. It does require code execution on the machine driving the display. The Light Auditor defense (2022) was built to address the Light Ears infrared channel via power monitoring. LightComm's power variation at swing=2 is well below Light Auditor's detection resolution.

Light Ears (Maiti and Jadliwala, 2019) is the most direct predecessor and uses the same device class, the same unauthenticated UDP protocol, and the same network-only threat model. The contribution here is demonstrating that visible-spectrum operation is achievable through perceptual color science and adaptive receiver design.

## 6.2 Throughput

The effective data rate for the color axis channel is 0.24 data bit/s. For dimming OOK it is 0.44 data bit/s at the same 750 ms symbol period. The difference arises from the lower overhead of 3x repetition coding relative to Hamming (7,4). Both rates are limited by the WiZ 8285 firmware's 500 ms smoothing floor. The channel physics does not set them.

Three factors account for the throughput, in order of magnitude. The 500 ms smoothing floor caps the raw symbol rate at 1.33 Hz. Calibration overhead consumes 50% of wire bits. FEC consumes a further 43% for Hamming(7,4) or 67% for 3x repetition. Removing the smoothing floor by substituting a WLED-compatible LED controller would raise the symbol rate to 60 Hz or more. At 60 Hz with 50% CAL overhead and 3x repetition coding, the effective payload rate would be approximately 10 data bit/s, comparable to the aIR-Jumper infrared channel. Figure 6 shows the wire-frame composition and the efficiency cascade.

### 6.3 Detectability

The setPilot commands at 750 ms intervals are not distinguishable from a user cycling through lighting presets via the WiZ app. Standard IoT monitoring approaches are blind to photonic output content by design.

Temporal variance analysis of camera video pointed at the lamp is the one approach that could detect the channel. A LightComm transmission at 750 ms per bit produces a component at 667 mHz in the camera ROI time series, visible in the frequency spectrum of the mean luminance or R-channel signal. Implementing this detection requires security camera analytics with field of view covering the lamp and sufficient temporal resolution to resolve sub-Hz periodicity. No such capability is present in currently deployed smart home or enterprise IoT security tools.

### 6.4 Limitations

*Table 6. Study limitations and proposed next steps.*

| Limitation | Root cause | Proposed next step |
|---|---|---|
| **Throughput 0.24 bit/s (color axis)** | WiZ 8285 PWM smoothing (~500 ms) | Test with WLED-compatible controller; 60 Hz symbol rate achievable |
| **Color axis gamut clipping on warm bulbs** | 2700K bulb clips commands outside warm gamut | Use L* brightness axis or dimming OOK for warm bulbs |
| **All results are single or dual runs** | No statistical replication | Run n >= 5 per condition; report mean and standard deviation |
| **AWB suppression incomplete** | DSHOW API varies by firmware | Test hardware-register AWB lock on machine-vision cameras |
| **Range and angle uncharacterized** | Camera inside lampshade only | Characterize signal at 1 to 5 m with telephoto zoom |
| **Smart plug channel unvalidated** | Concept only | Acquire WiZ plug; repeat on/off protocol with conventional lamp |
| **SIEM invisibility unverified** | Theoretical claim | Record matched session pairs; submit to SIEM rule set |
| **Human luminance JND not empirically tested** | Perceptual claim not verified by observer study | Run n >= 10 ABX study at swing=2 and swing=6 |

**Table 6** lists known limitations with root causes and proposed next steps. The most consequential gap is the absence of repeated measurements. Every row in Tables 3 and 5 is based on one or two runs. The claimed 96.2% wire accuracy at swing=2 may be stable or may reflect favorable sampling from a wider distribution. Statistical characterization requires n >= 5 runs per condition with mean and standard deviation reported.

The color axis channel showed gamut clipping on the 2700K warm bulb. Commanding RGB (196, 184, 224) is outside the warm bulb's achievable color space. Both symbol colors clip to the same output and the observed separation collapses near zero regardless of JND. The dimming OOK channel is not affected by this because brightness modulation works on any bulb. Future work on the color axis channel should use the L* brightness axis instead of a* for warm bulbs, since L* modulation scales all channels proportionally and stays within gamut across all color temperatures.

The perceptual claim for swing=2 dimming OOK requires qualification. The human brightness JND under photopic conditions is approximately 1 to 2 percent of adapting luminance. At center dimming=55 with swing=2, the relative change is approximately 3.6 percent, which approaches the JND threshold without falling below it. Substantiating the imperceptibility claim requires a formal ABX observer study under the experimental lighting conditions. The current evidence shows that swing=2 passes the decode test. It does not show that swing=2 is undetectable by human observers.

### 6.5 Future Work on Axis Selection and Real-Time Optimization

The closed-loop optimizer described in Section 4.6 currently evaluates only one modulation axis at a time and uses a sequential swing search. A more efficient approach would model the separation-versus-swing curve from a small number of samples and predict the minimum viable swing without exhaustive search, reducing setup time from approximately 20 minutes to under 3 minutes.

Automatic axis selection is the other opportunity. The optimal modulation axis depends on the specific bulb's spectral output, the camera's channel sensitivity, the lampshade's chromatic transfer function, and the ambient illumination. The current code selects a* by default and switches to dimming OOK manually when gamut clipping is observed. The optimizer could instead measure the achieved separation on a*, b*, L*, and dimming simultaneously during the calibration phase and select the axis with the best observed-to-commanded separation ratio. On the 2700K warm bulb used in the dimming experiments, this would have automatically selected brightness modulation and avoided the extended debugging session that identified the gamut clipping problem.

The browser dashboard confirmed the gamut clipping problem interactively. The live sparkline showed near-zero separation despite increasing JND, directing attention to the base color mismatch. Extending the dashboard to display the optimizer's per-axis separation measurements and its predicted operating envelope would provide the operator with a quantitative picture of the tradeoff between stealth and reliability in current ambient conditions.

## 7. Conclusion

Maiti and Jadliwala (2019) demonstrated smart bulb infrared exfiltration and declined to use visible light because color changes would be noticed. This paper shows that visible-spectrum operation is achievable through perceptual color science and adaptive receiver design. Two channels are validated. The color axis channel operates at JND = 0.5, below the human color discrimination threshold. The dimming OOK channel operates at a 2-unit brightness swing that is at the edge of the human luminance threshold under typical indoor conditions.

The color axis channel required an adaptive receiver because AWB drift is larger than the signal at sub-JND amplitudes. Interleaved pilot symbols allow the receiver to track the drift instead of holding a fixed reference. The dimming OOK channel avoids the AWB problem and also avoids the gamut clipping that limits the color axis channel on warm bulbs. It is the more practical choice for bulbs with color temperatures below approximately 3500K.

A closed-loop optimizer watches the bulb through the camera, measures observed separation in real time, and finds the minimum viable operating point without manual calibration. This was used during the experiments to diagnose the gamut clipping problem on the warm bulb and redirect the investigation to the dimming channel.

Neither channel is detectable by network monitoring, power consumption analysis, or visual inspection. Detection requires temporal variance analysis of camera video aimed at the lamp. The six additional WiZ protocol modalities surveyed in Section 3 represent a broader channel surface, and the smart plug variant transfers the on/off channel to any conventional lamp on a WiZ outlet without changes to the transmitter code.

**Acknowledgements.** The author thanks the PeopleTec Technical Fellows Program for their encouragement and research support.

## References

**[1]** Loughry, J. and Umphress, D. A. (2002). Information leakage from optical emanations. *ACM Transactions on Information and System Security, 5*(3), 262-289. doi:10.1145/545186.545189.
**[2]** Loughry, J. (2018). Optical TEMPEST. *Proc. EMC Europe 2018, 172-177. doi:10.1109/EMCEurope.2018.8485128.*
**[3]** Guri, M., Zadov, B., Daidakulov, A., and Elovici, Y. (2018). xLED: Covert data exfiltration from air-gapped networks via switch and router LEDs. Proc. PST 2018, 1-12 (arXiv:1706.01140).
**[4]** Guri, M., Zadov, B., Atias, E., and Elovici, Y. (2017). LED-it-GO: Leaking data from air-gapped computers via the hard drive LED. Proc. DIMVA 2017, 161-184.
**[5]** Guri, M., Bykhovsky, D., and Elovici, Y. (2019). aIR-Jumper: Covert air-gap exfiltration via security cameras and IR. *Computers and Security, 82, 15-29. doi:10.1016/j.cose.2018.11.004 (arXiv:1709.05742, 2017).*
**[6]** Guri, M., Bykhovsky, D., and Elovici, Y. (2020). BRIGHTNESS: Leaking sensitive data from air-gapped workstations via screen brightness. *arXiv:2002.01078*
**[7]** Guri, M., Zadov, B., Bykhovsky, D., and Elovici, Y. (2019). CTRL-ALT-LED: Leaking data from air-gapped computers via keyboard LEDs. Proc. IEEE COMPSAC 2019, 801-810.
**[8]** Maiti, A. and Jadliwala, M. (2019). Light Ears: Information leakage via smart lights. *Proc. ACM IMWUT, 3*(3), Article 98. doi:10.1145/3351256.
**[9]** Jung, W., Cui, K., Koltermann, K., Wang, J., Xin, C., and Zhou, G. (2022). Light Auditor: Power measurement can tell private data leakage through IoT covert channels. Proc. ACM SenSys 2022, 518-532. doi:10.1145/3560905.3568535.
**[10]** Zhou, A., Su, G., Zhu, S., and Ma, H. (2019). Invisible QR code hijacking using smart LED. Proc. ACM IMWUT, 3(3). doi:10.1145/3351284.
**[11]** Ronen, E., Shamir, A., Weingarten, A., and O'Flynn, C. (2017). IoT goes nuclear: Creating a ZigBee chain reaction. Proc. 2017 IEEE Symposium on Security and Privacy, 195-212. doi:10.1109/SP.2017.14.
**[12]** WiZ Connected Lighting Shanghai Co., Ltd. (2019). WiZ 8285 1517PT User Manual V19.09.1. FCC ID 2AUOF-1517PT.
**[13]** SVPRO (2024). 2MP Global Shutter USB Camera, 5-50 mm telephoto lens. Product specification.
**[14]** Sharma, P. and Awasthi, L. K. (2024). Unveiling the hidden dangers: Security risks and forensic analysis of smart bulbs. *Forensic Science International: Digital Investigation, 50, 301794. doi:10.1016/j.fsidi.2024.301794.*
**[15]** Rayhan, A. F., Riskiana, R. R., Kusumo, D. S., Wibowo, R. O. S., and Praja, M. R. A. (2024). Model-based testing of IoT mobile controller smart light systems using extended finite state machine. Proc. 12th ICoICT 2024, 123-130. IEEE. doi:10.1109/ICoICT61617.2024.10698163.
**[16]** Panwar, J. and Rohilla, R. (2024). Security concerns in IoT light bulbs: Investigating covert channels. Computer Science and Information Technology (CS and IT), 14(15), 91-99. doi:10.5121/csit.2024.141508.

**[17]** Zhang, K., Wu, C., Yang, C., Zhao, Y., Huang, K., Peng, C., Liu, Y., and Yang, Z. (2018). ChromaCode: A fully imperceptible screen-camera communication system. Proc. ACM MobiCom 2018, 575-590. doi:10.1145/3241539.3241543.
**[18]** Cronin, P., Gouert, C., Mouris, D., Tsoutsos, N. G., and Yang, C. (2019). Covert data exfiltration using light and power channels. Proc. 37th IEEE International Conference on Computer Design (ICCD), 301-304.

## Appendix A. Protocol

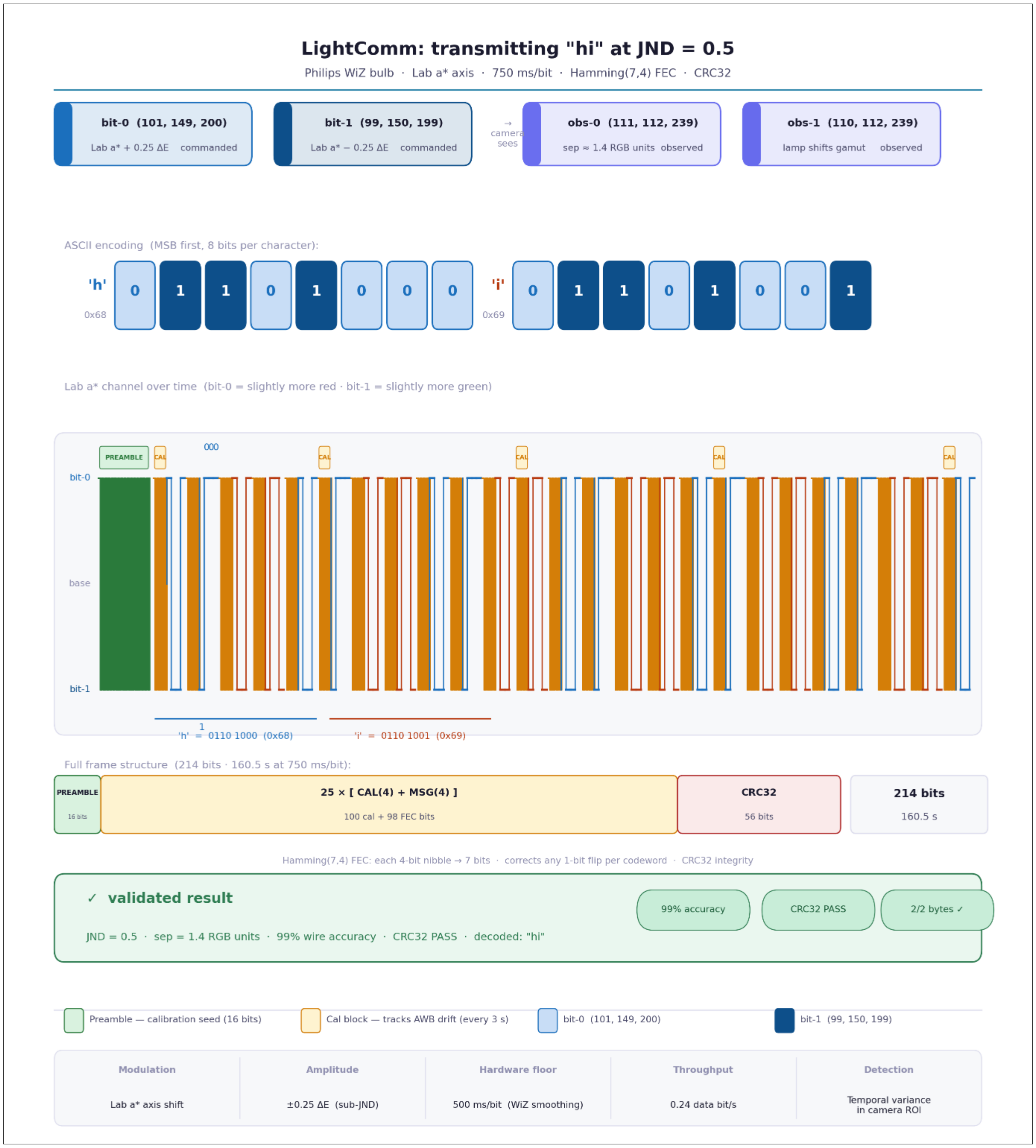

*Figure 9. Frame structure for 'hi' at JND = 0.5. Top: commanded and observed color swatches. Center: bit encoding and Lab a* waveform with PREAMBLE (green) and CAL blocks (amber). Bottom: frame anatomy (214 wire bits, 160.5 s) and validated result.*